\documentclass[twocolumn]{aastex62}
\usepackage{multirow}

\graphicspath{{./}{figures/}}

\received{January 1, 2018}
\revised{January 7, 2018}
\accepted{\today}
\submitjournal{ApJ}

\shorttitle{Old and dim: PSR J1154--6250}
\shortauthors{Igoshev et al.}

\begin{document}

\title{Discovery of X-rays from the old and faint pulsar  J1154--6250}

\correspondingauthor{Andrei P. Igoshev}
\email{ignotur@gmail.com, ignotur@physics.technion.ac.il}

\author[0000-0003-2145-1022]{Andrei P. Igoshev}
\affil{Department of Physics, Technion – Israel institute of Technology,
Haifa 3200003, Israel \\
}

\author{Sergey S. Tsygankov}
\affiliation{Department of Physics and Astronomy, FI-20014 University of Turku, Finland\\}
\affiliation{Space Research Institute of the Russian Academy of Sciences, Profsoyuznaya Str. 84/32, Moscow 117997, Russia}

\author{Michela Rigoselli}
\affiliation{INAF, Istituto di Astrofisica Spaziale e Fisica Cosmica Milano, via E. Bassini 15, I-20133 Milano, Italy}
\affiliation{Dipartimento di Fisica G. Occhialini, Universit\`a degli Studi di Milano Bicocca, Piazza della Scienza 3, I-20126 Milano, Italy}

\author{Sandro Mereghetti}
\affiliation{INAF, Istituto di Astrofisica Spaziale e Fisica Cosmica Milano, via E. Bassini 15, I-20133 Milano, Italy}

\author{Sergei B. Popov}
\affiliation{Sternberg Astronomical Institute, Lomonosov Moscow State University, Universitetsky prospekt 13, 119234, Moscow, Russia}

\author{Justin G. Elfritz}
\affiliation{Anton Pannekoek Institute, University of Amsterdam, Science Park 904, 1098 XH Amsterdam, The Netherlands}

\author{Alexander A. Mushtukov}
\affiliation{Anton Pannekoek Institute, University of Amsterdam, Science Park 904, 1098 XH Amsterdam, The Netherlands}
\affiliation{Space Research Institute of the Russian Academy of Sciences, Profsoyuznaya Str. 84/32, Moscow 117997, Russia}
\affiliation{Pulkovo Observatory, Russian Academy of Sciences, Saint Petersburg 196140, Russia}

\begin{abstract}

 We report on the first X-ray observation of the 0.28 s isolated radio pulsar PSR J1154--6250 obtained with the {\it XMM-Newton} observatory in February 2018. A point-like source is firmly detected at a position consistent with that 
 of PSR J1154--6250. The two closest stars are outside the 3$\sigma$ confidence limits of the source position and thus unlikely to be responsible for the observed X-ray emission. 
 The energy spectrum of the source can be fitted equally well either with an absorbed power-law with a steep photon index $\Gamma\approx 3.3$ 
or with an absorbed blackbody with temperature $kT=0.21\pm 0.04$~keV and emitting radius $R_\mathrm{BB} \approx 80$ m  
(assuming a distance of 1.36~kpc). 
The X-ray luminosity of  $4.4\times 10^{30}$ erg s$^{-1}$ derived with the  power-law fit corresponds to an efficiency of
$\eta_X = L^\mathrm{unabs}_X/\dot E = 4.5\times 10^{-3}$,  similar to those of other  old pulsars.  The X-ray properties of PSR J1154--6250 are consistent with an old age and suggest that the  spatial coincidence of this pulsar with the  OB association  Cru OB1 is due to a chance alignment. 

\end{abstract}


\section{Introduction}

Rotation-powered pulsars were first discovered as radio sources emitting pulses at extremely precise  periodic  intervals 
\citep{1968Natur.217..709H}, 
and later detected also in the optical \citep{1969Natur.221..525C}, X-ray \citep{1992Natur.357..222H,1993Natur.361..136O} and $\gamma$-ray bands. 
Their properties   are   well explained by a model of isolated rotating neutron star (NS) with magnetic field $B\sim 10^{12}$~G \citep{1969Natur.221...25G, RS}.
From the rotation period $P$ and its time derivative $\dot P$, one can define a characteristic   age $\tau=P/(2\dot P)$ and measure the rotational energy loss rate $\dot E = -4\pi^2I\dot P / P^3$, where $I$ is the NS moment of inertia. 

The X-ray properties of isolated radio pulsars (see \citealt{review} for a  review) depend on their age.
The youngest, $\tau < 10^4$ years, and energetic, $\dot E > 10^{36}$~erg s$^{-1}$, objects emit mainly non-thermal radiation, with strong pulsations and a spectrum well described by a  power-law model. At intermediate stages, $10^4<\tau<10^6$~years, when the NS is still hot, pulsars show a combination of  magnetospheric non-thermal emission and thermal emission from the whole surface, the latter typically described, as a first approximation, by a blackbody model. 
Without additional heating mechanisms, unlike the magnetar case, a NS typically reaches a thermal luminosity of $\sim 10^{30}$~erg s$^{-1}$ in $\lesssim1$~Myr \citep{2004ApJS..155..623P}. At characteristic ages larger than  $\approx 1$~Myr, the NS cools down in absence of heating sources \citep{ns_cooling} and its surface radiation becomes undetectable for the currently available X-ray telescopes. 

However, some of these old objects show thermal emission with temperature $kT\approx 0.1-0.3$~keV and emission radius $R_\mathrm{BB}\approx 10-200$~m \citep[see e.g.,][]{2004ApJ...616..452Z,2008ApJ...685.1129M,2008ApJ...686..497G,2012Sci...337..946K,2016ApJ...831...21M, 2017MNRAS.466.1688H,rigoselli2018}. 
The thermal emission in these old pulsars is  interpreted as   radiation from a polar cap heated by    particles accelerated in the magnetosphere \citep{harding_muslimov1,hardingmuslimov2}. The size of the cap, as estimated from blackbody fits to the  X-ray spectra, is often smaller than the size  predicted by theory in the case of a dipolar field, but fitting the spectra with   NS atmosphere models \citep{2004ApJ...616..452Z} yields in some cases emission regions with a size consistent with the predicted values. If old pulsars show only non-thermal emission, the X-ray efficiency (the ratio of the X-ray luminosity to the spin-down energy loss rate) often exceeds the values seen in  younger pulsars \citep{2006ApJ...636..406K}. The reason for this is unknown, but  it could be partly explained by a selection effect. So, a larger sample and longer exposures are needed to gain a better understanding of the X-ray emission of old radio pulsars \citep{2012ApJ...749..146P}.

It must also be considered that the spin-down age $\tau$,  which is used to interpret the X-ray properties,   is a reasonable estimate of the true age of a pulsar only if two conditions are met: (1) the initial rotational period  is much smaller than its current value and (2) the braking index $n=2 - P\ddot P / \dot P^2 \equiv 3$. From observations it is known that, in some cases,  the initial period can be quite long and comparable with the observed period \citep{2012Ap&SS.341..457P}. The observed braking indexes range from 0.9(2) for PSR J1734--3333 to 3.15(3) for PSR J1640--4631 \citep{high_braking_index, 2011ApJ...741L..13E}. The difference with respect to the expected value $n$ = 3 
could be caused by deviation from counter-alignment between the rotational axis and the orientation of the magnetic dipole \citep{2014MNRAS.441.1879P} or by  some evolution of the magnetic field.

\begin{deluxetable}{ll}
\tablecaption{Observed and derived parameters for  PSR J1154--6250\label{t:psr_prop}}
\tablehead{
\colhead{Parameter} & \colhead{Value} 
}
\startdata
R.A. (J2000.0)\dotfill                                               & 11$^\mathrm{h}$54$^\mathrm{m}$20$^\mathrm{s}$.1(1) \\
Dec. (J2000.0)\dotfill                                               & --62$^\circ$50$'$02$\farcs$7(7) \\
Period $P$ (s)\dotfill                                     & 0.28201171065(3)\\
Period derivative $\dot P$ (s s$^{-1}$)\dotfill                         & $5.59(5)\times 10^{-16}$ \\
Spin-down age $\tau$ (years)\dotfill                       & $8\times 10^6$  \\
Surface dipolar magnetic field $B$ (G)\dotfill               & $4.0\times 10^{11}$\\
Spin-down energy loss rate $\dot E$ (erg s$^{-1}$)\dotfill & $9.8\times 10^{32}$ \\
Dispersion measure DM (cm$^{-3}$ pc)\dotfill               & 74(6)\\
Distance based on DM (kpc)\dotfill                         & 1.36\tablenotemark{a}\\
\enddata
\tablenotetext{a}{Distance is derived using the electron density model of \cite{2017ApJ...835...29Y}}
\tablecomments{Information is based on \cite{obser_properties} and the ATNF pulsar catalogue. Numbers in parentheses show the uncertainty for the last digits. \\}
\end{deluxetable}

Here we report on a recent X-ray observation of the rotation powered pulsar PSR J1154--6250 obtained with the {\it XMM-Newton} satellite.
The main parameters of PSR J1154--6250 are summarized in Table~\ref{t:psr_prop}, based on the information in the ATNF pulsar catalogue v.1.58\footnote{ http://www.atnf.csiro.au/research/pulsar/psrcat/} \citep{atnf}. The dispersion measure corresponds to a distance of 1.77 kpc according to the NE2001 electron density model  \citep{2002astro.ph..7156C} and to 1.36 kpc according to the newer model by \cite{2017ApJ...835...29Y}. In the following we use $d$ = 1.36 kpc.

PSR J1154--6250 is located at an angular distance of 
1.6 degrees from the center of  the OB association Cru OB1 \citep{ob_dist}, which has coordinates R.A.=11$^\mathrm{h}$40$^\mathrm{m}$00$^\mathrm{s}$.0 Dec.=--62$^\circ$54$'$00$\farcs$ A typical size of the association of 40~pc   translates to $1^\circ 9'$ at 2~kpc distance. Thus PSR J1154--6250 could be  associated with   Cru OB1, implying a true age much smaller than its characteristic  age.  NSs are known to receive large natal kicks at the moment of their formation \citep{1994Natur.369..127L,verbunt2017}, with average speed 370~km s$^{-1}$ which translates to a traveled distance $\approx 3$~kpc in 8~Myr. The natal kick makes these objects oscillate in the Galactic potential.  
Time interval of $\tau < 20$ Myr is not enough to finish a half of the oscillation in the potential, so NSs with the age in the range 1--20 Myr could be projected onto OB association only if they move along the Galactic disc, which is a quite rare situation.  Among NSs with such spin-down ages there could be objects with complicated magnetic field evolution, caused for example by the fall-back of matter after the supernova explosion \citep{fall_back} and subsequent magnetic field re-emergence \citep{2011MNRAS.414.2567H,2012MNRAS.425.2487V,igoshev2016}. So, the primary goal of this research is to check if PSR J1154--6250 is as old as it follows from its spin-down age.

 \section{Observations and data reductions}

The observations were performed on 8th February 2018 with the \textit{XMM-Newton} telescope, using the European Photon Imaging Camera (EPIC). All the EPIC cameras, the pn and the two MOS \citep{epn,mos}, were used in full-frame mode and with the thin optical  filter (see Table~\ref{t:journal}). The nominal exposure was 60 ks. The analysis was done using the \textit{XMM-Newton} Science Analysis System (SAS) package version 16.1.0.

Since the object is quite dim,
we performed the analysis with  two different methods adequate for faint sources: (1) a traditional spectral analysis, 
based on source and background extraction regions, but using 
the $W$-statistics for the spectral fits \citep{w_statistics},
and (2) a maximum likelihood (ML) method \citep{2017MNRAS.466.1688H,rigoselli2018}.

\subsection{Traditional analysis}

\begin{figure*}
        \centering
        \begin{minipage}{0.48\linewidth}
        \includegraphics[width=1\columnwidth]{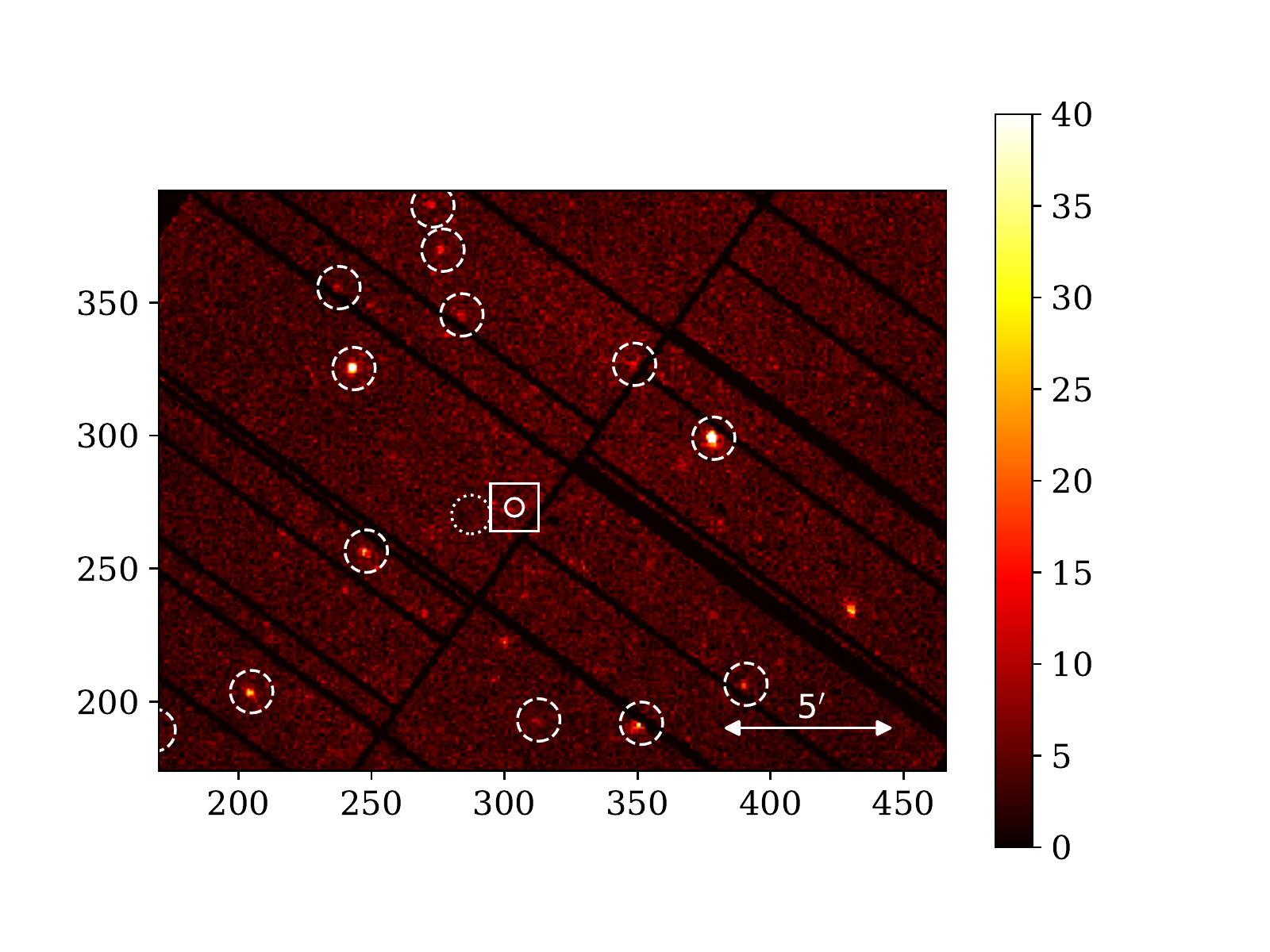}
        \end{minipage}
        \begin{minipage}{0.41\linewidth}
        \includegraphics[width=1\columnwidth]{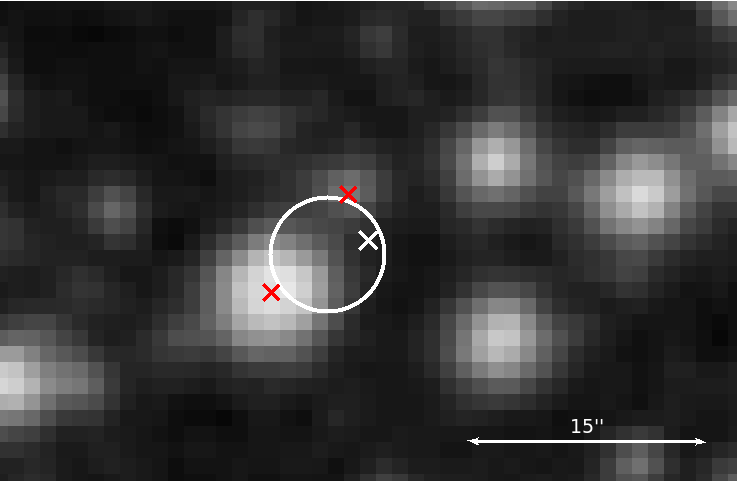}
        \end{minipage}
        \caption{Filtered EPIC pn image (energy range 0.3-2~keV) of the field  binned into 4 pixels per bin (left panel). The white rectangle region is scaled and shown in the right panel based on the DSS2 survey (emulsion and filter IIIaF+OG590; bandpass peaks at around 6500 \AA \hspace{0.05cm} which corresponds to red color). White solid circles show the spectral extraction region used in the traditional approach (left panel) and $3\sigma$ confidence interval for coordinates of the X-ray source (right panel). White dashed circles are locations of sources used to improve astrometry in the field. White dotted circle shows the location of the background extraction region used in the traditional approach. Red crosses show exact positions of the closest stars in the second Gaia data release. White cross is for radio position of the PSR J1154-6250. The colorbar shows number of counts per bin. } 
\label{f:image}
\end{figure*}

Using the source detection pipeline of the SAS, we detected with a likelihood of 36
a point source  with $81\pm 15$ pn counts (0.2-12~keV energy range) close to the pulsar position. 
We improved the \textit{XMM-Newton} astrometry using the  second Gaia data release, as described in the Appendix. The corrected source coordinates are
R.A.=11$^\mathrm{h}$54$^\mathrm{m}$20$^\mathrm{s}$.52 and Dec.=--62$^\circ$50$'$03$\farcs$6 (J2000.0),  with an error of  $1\farcs 2 $. 
This position differs by  3$''$ (corresponding to 2.5$\sigma$)
from that of PSR J1154--6250,  determined by means of timing radio observations (see Table~\ref{t:psr_prop} and Figure~\ref{f:image}). We note that the closest stars to the X-ray positions are at more than 4$''$, i.e. outside the 3$\sigma$ error radius. 
We consider unlikely that one of them could be the counterpart of the detected X-ray source, which we therefore identify with PSR J1154--6250. 

Due to the limited flux from the source, we analyzed events recorded by the three cameras of the EPIC instrument: both MOS and pn. We created good time intervals (GTI) excluding background flare periods (count rates exceeding $2.46$, $3.43$,   and 9.5 counts s$^{-1}$ for MOS1, MOS2 and pn, respectively) 
identified by the procedure of S/N optimization described in \cite{2016A&A...590A...1R}.
After GTI filtering, we kept only the events   in the range 0.2-10~keV which satisfy standard pattern requirements (\texttt{PATTERN} $\leq 12$ for MOS and \texttt{PATTERN} $\leq 4$ for pn) and quality flags.

The source spectrum was extracted from a circle  centered at the radio coordinates and with   radius 15 arcsec. The background spectrum was extracted from a circle centered at R.A.=11$^\mathrm{h}$54$^\mathrm{m}$30$^\mathrm{s}$.6 and Dec.=--62$^\circ$50$'$14$\farcs$8 with radius 32 arcsec. The auxiliary and response files for the spectral analysis were prepared using the standard tasks \texttt{rmfgen} and \texttt{arfgen}. We combined the   MOS1 and MOS2 spectra into a single file using \texttt{epicspeccombine} and analyzed them using averaged response matrix. 

We fitted simultaneously the pn and MOS  spectra using version 12.9.1 of the \texttt{xspec} software.
For the interstellar absorption we adopted  the \texttt{tbabs} model which is based on photoionization cross-sections by \cite{2000ApJ...542..914W}. 

Due to the small count rate we rebinned the spectra to have one  count per energy bin and used the W-statistics (\texttt{cstat} option in \texttt{xspec} \citealt{w_statistics}), which is a version of  C-statistics \citep{cash1979} applied if no model for the background radiation is provided. A comparison of different models was done by means of the Akaike information criterion (AIC, see \citealt{1974ITAC...19..716A}), which is used in astronomy \citep{2007MNRAS.377L..74L} and in particular in X-ray astronomy to discriminate between different spectral models   \citep{2017MNRAS.470..126T}. The value of AIC is computed as $\mathrm{AIC = 2k+C}$, where C is the minimum value of statistics   and $k$   the number of model parameters. A difference of 10 between the AIC values computed for two models is significant to choose the best model and corresponds to a case when one model is $\exp(10/2)\approx 150$ times more probable than the other.

\begin{table*}
\caption{Details of the \textit{XMM-Newton} observation}
\label{t:journal}
\begin{center}
\begin{tabular}{ccccccc}
\hline
\hline
Pulsar         & Obs. ID    & Start time & End time & \multicolumn{3}{c}{Effective exposure (s)} \\
               &            & UT         &  UT      & pn & MOS1 & MOS2 \\
\hline
PSR J1154--6250 & 0804240201 & 2018 Feb 08 01:11:59 & 2018 Feb 08 18:10:02 & 49950 & 58900 & 58900 \\
\hline
\end{tabular}
\end{center}
\end{table*}

\subsection{Maximum likelihood analysis}

The ML method estimates the most probable number of source and background counts by comparing the spatial distribution of the observed counts with the expected distribution of a model in which there is a source
and a uniform background.
The expected spatial distribution of the counts for a point source is given by the instrument point spread function (PSF). 
We used the MOS and pn PSF
derived from the {\it XMM-Newton} in-flight calibrations\footnote{http://www.cosmos.esa.int/web/xmm-newton/calibration/ documentation}, with parameters appropriate for the average energy value in each of the considered bins.

We used single- and multiple-pixel events for both the pn and MOS. The events detected in the two MOS cameras were combined into a single data set, and we used averaged response files.

Since with the  maximum likelihood analysis  it is possible to derive an accurate estimate of the background at the source position, it was not necessary to remove the time intervals of high background. We applied the ML  in a circular region positioned to avoid the gaps between the CCD chips (radius of $60\farcs$,  center at R.A.=11$^\mathrm{h}$54$^\mathrm{m}$23$^\mathrm{s}$.2,  Dec.=--62$^\circ$49$'$47$''$). 
For the spectral extraction  we fixed the source position at the coordinates found by the ML in the pn+MOS image in the $0.4-2$ keV range
(R.A.=11$^\mathrm{h}$54$^\mathrm{m}$20$^\mathrm{s}$.4, Dec.=--62$^\circ$50$'$04$\farcs$4).
This position is consistent with that found with the SAS pipeline.

In the energy range $0.4 - 2$ keV the source was detected with $113 \pm 21$ counts (pn) and $66 \pm 15$ counts (MOS), which correspond to a total count rate of $(3.4 \pm 0.5) \times 10^{-3}$ count s$^{-1}$.

The energy bins for the spectral analysis were chosen from the requirement to have a signal-to-noise ratio greater than $2.5 \sigma$ in each spectral channel. This resulted in four energy bins for the pn and   three     for the MOS. Below 0.4 keV and above 2 keV, the source detection was below our significance threshold; therefore we derived upper limits on the source counts in these energy bins.

\section{Possible optical counterparts}

 In the Gaia database there are no stars closer than $4''$ to the corrected X-ray position, see Figure~\ref{f:image}. The closest object, at $4\farcs 06$,  is Gaia DR 2 5334588220496359424 which has $g=17.48$.
A slightly brighter star, Gaia DR2 5334588151811060992 with $g=14.15$, is located at  $4\farcs 29$. 
It has $B=16.4$, $I=13.11$, $J=11.937$ and $K=10.951$, 
while the fainter source has $B=19.6$, $I=16.08$, $J=14.07$ and $K=12.86$ \citep{denis}. Based on the color differences $I-J$ and $J-K$, we identify the brighter star as a main sequence of spectral type K5 and the fainter one  as an M5 star.

We fitted the  spectra of our X-ray source with thermal plasma models  (\texttt{apec} by \citealt{apec}, \texttt{mekal} by \citealt{mekal} ),
as expected in the case of coronal emission from normal stars.  
Acceptable fits could be obtained with temperatures in the range  $kT=0.23-0.26$ keV,  
absorption $N_H$$\sim$ 10$^{22}$~cm$^{-2}$,  and flux $\sim$3$\times$10$^{-15}$ erg cm$^{-2}$ s$^{-1}$. 
The resulting X-ray to optical flux ratios  $\log(f_x/f_J) = \log(f_x) + 0.4 J +6.30$ \citep{agueros} are $-3.34$  and $-2.6$ for the K and M star, respectively. 
These values are compatible with X-ray emission from late type stars and therefore we cannot exclude these objects as potential counterparts of the X-ray source based on spectral and/or flux arguments. However, we note that both stars are outside than 3$\sigma$ confidence level error circle of the X-ray source and therefore we consider that PSR J1154--6250 is a much more likely responsible for the detected X-ray emission.

\section{Results}

\begin{figure}
        \centering
        \includegraphics[width=1\columnwidth]{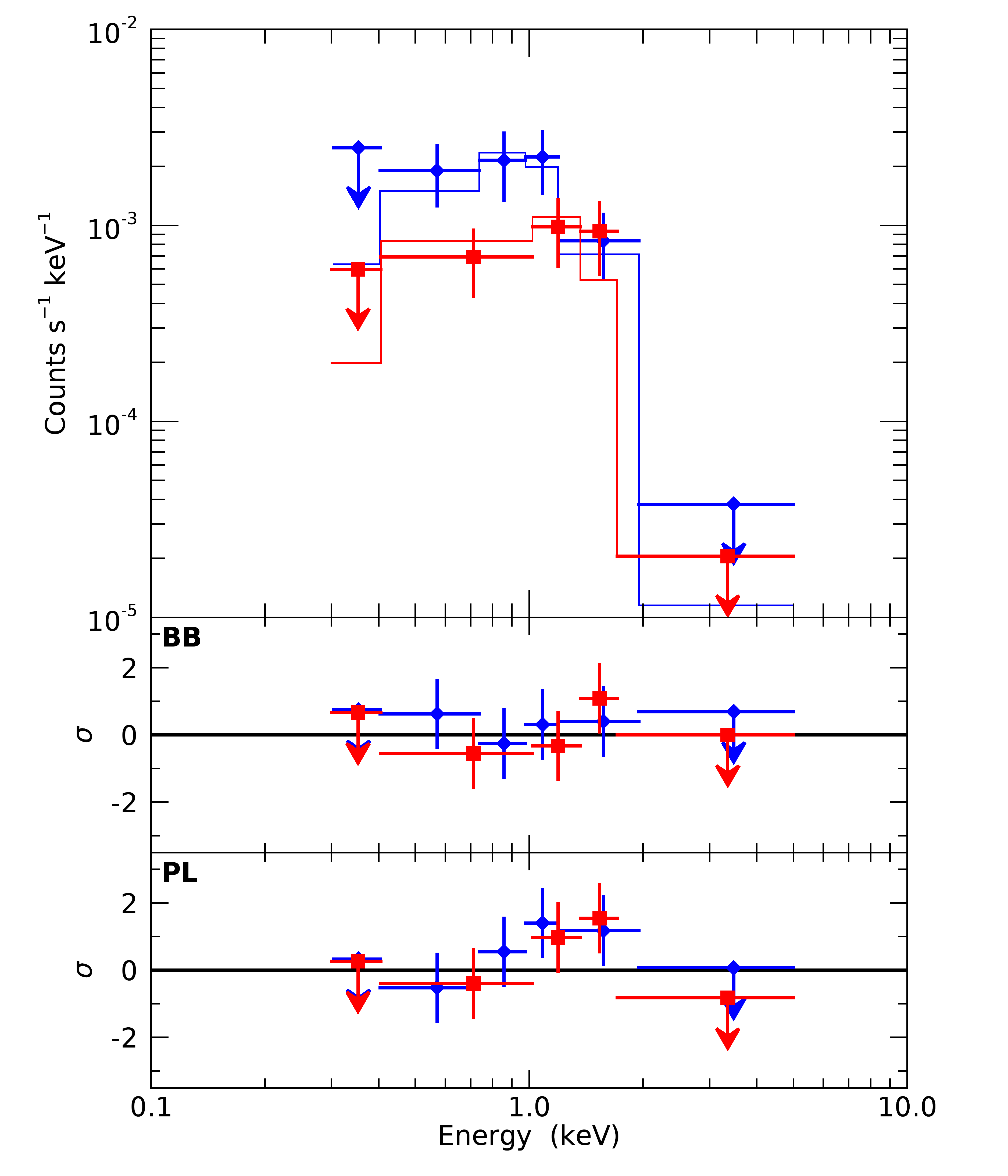}
        \caption{EPIC pn (blue diamonds) and MOS (red squares) X-ray spectra of PSR J1154--6250 extracted with the maximum likelihood method. The best fit blackbody model is shown in the top panel, and the corresponding residuals in the middle panel. The bottom panel shows the residuals obtained by fitting the spectra with a power-law.} 
\label{f:MLspe}
\end{figure}

\begin{figure}
\includegraphics[width=1\columnwidth]{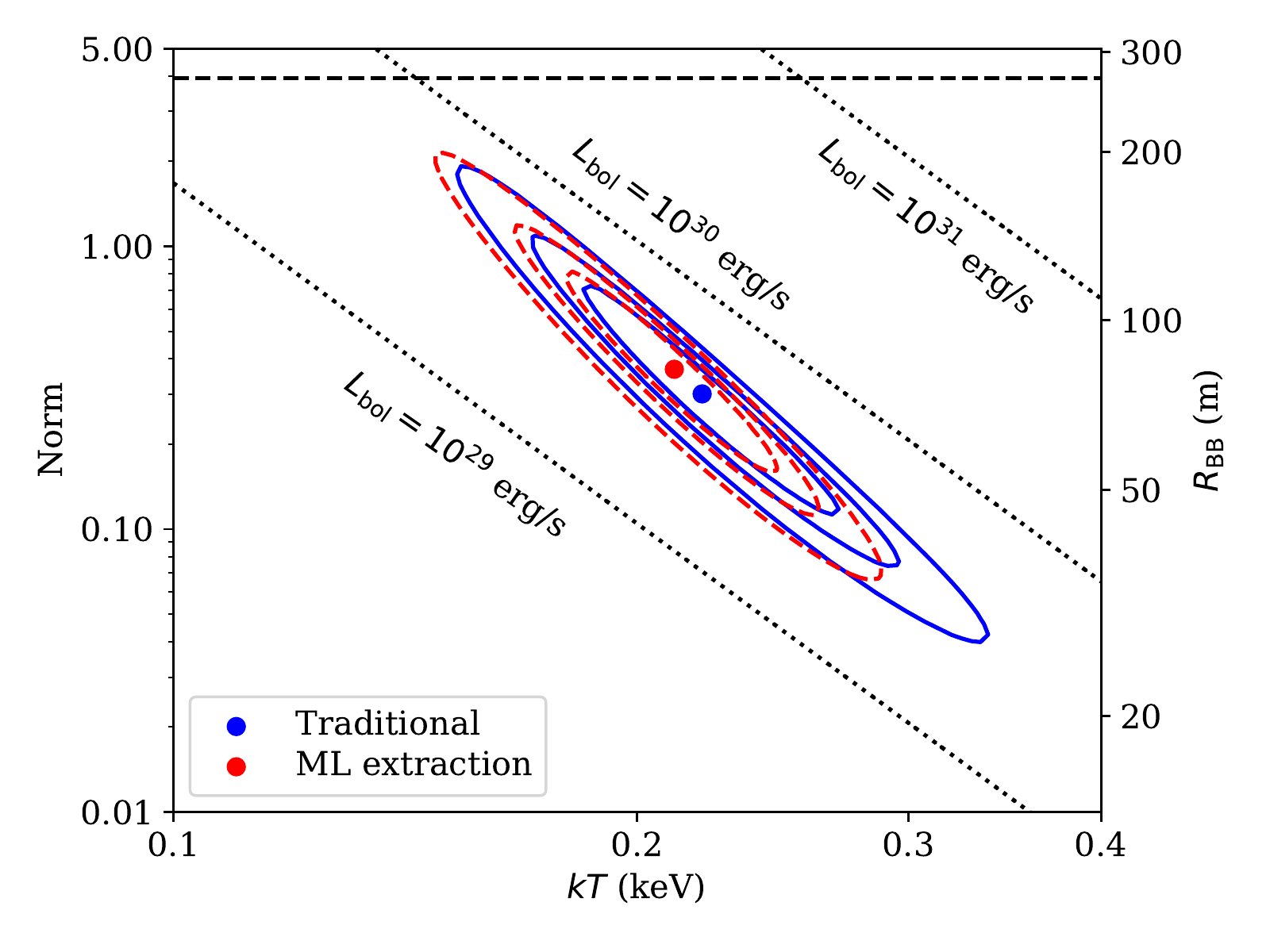}
\caption{The most probable value (dot) and confidence intervals (1,2 and 3 $\sigma$) of temperature and emission radius for the   blackbody model. Traditional analysis are the solid blue lines, while  the ML technique are the dashed red lines. The dashed horizontal line shows the size of the dipolar polar cap calculated using Eq.\ref{e:Rpc} and assuming a NS radius of 10 km. The normalization is computed as $R_{km}^2/D^2_{10}$ where $R_\mathrm{km}$ is the emission radius in km and $D_{10}$ is the distance to the source in units of 10 kpc.}
\label{f:correlation}
\end{figure}

The results of the spectral analysis obtained with the two methods are summarized in Table~\ref{t:res}. The absorption column density was  poorly constrained due to the small count rate. So we decided to fix it  at $N_\mathrm{H} = 2.2\times 10^{21}$~cm$^{-2}$,  which follows from the relation between dispersion measure and $N_\mathrm{H}$ for radio pulsars \citep{nh_dm}. 

With the traditional analysis the absorbed blackbody model fits the data slightly better, however the difference of 3.68 in the AIC values is not enough to prefer this model  to the power-law. 
The maximum likelihood analysis provides results in full agreement with those of the traditional analysis. 
Also in this case, both the blackbody and power-law models are acceptable (see Figure~\ref{f:MLspe}).

With a power-law  model we find  a rather high photon index value  $\Gamma\approx 3.3$,
but still compatible with the range of $\Gamma=2-4$ noticed by \cite{2012ApJ...749..146P}. The unabsorbed flux is $F^\mathrm{unabs}_{0.2-10~\mathrm{keV}}=(2.6^{+1.6}_{-1.0})\times 10^{-14}$~erg~s$^{-1}$~cm$^{-2}$.   

For the blackbody model we obtain a temperature $kT = 0.21 \pm 0.04$ keV and a radius of the emitting region $R_\mathrm{BB}=81_{-29}^{+46}$~m, which corresponds to a bolometric luminosity $L_\mathrm{bol} = \pi R_\mathrm{BB}^2 \sigma_B T^4 = 4\times 10^{29}$~erg s$^{-1}$, where $\sigma_B$ is the Stefan-Boltzmann constant. The radius and temperature anticorrelate with each other,  as shown in Figure~\ref{f:correlation}. The unabsorbed flux is $7.5_{-2.0}^{+2.2}\times 10^{-15}$~erg~s$^{-1}$~cm$^{-2}$. The values of $kT$ and $R_\mathrm{BB}$ found in our fit are similar to those found for old radio pulsars with thermal emission \citep[see e.g.,][]{2008ApJ...685.1129M,2016ApJ...831...21M}. 

We try also fits with the \texttt{NSA} model for   thermal emission from a NS hydrogen atmosphere  \citep{nsa1} with fixed parameters NS mass $M_\mathrm{NS}=1.4~M_\odot$ and radius $R_\mathrm{NS}=10$~km and $N_\mathrm{H} = 2.2\times  10^{21}$~cm$^{-2}$. This model gives $\log T_\mathrm{eff} \mathrm{[K]} = 6.15\pm 0.15$, $R_\mathrm{BB} = 350$~m, 
with C-value/d.o.f. = 197.02/212. The \texttt{NSA} model might not be the best choice since the magnetic field of PSR J1154--6250 is $4\times 10^{11}$~G which is not covered by this model.

\begin{table*}
\caption{Results of the spectral analysis. The best-fit values and 90\% confidence limits.}
\label{t:res}
\begin{center}
\begin{tabular}{lccccccccccccc}
\hline
\hline
\multirow{2}{*}{Model} & $N_\mathrm{H}$                & $\Gamma$ &  Norm.  & $kT$ & $R_\mathrm{BB}\tablenotemark{b}$  & $F^\mathrm{abs}_{0.2-10\;\mathrm{keV}}$  & $F^\mathrm{unabs}_{0.2-10\;\mathrm{keV}}$  & \multirow{2}{*}{Statistics}  \\
		               & $10^{22}$~cm$^{-2}$  & &\tablenotemark{a} & keV  & m                &  $10^{-15}$ erg s$^{-1}$ cm$^{-2}$ &  $10^{-15}$ erg s$^{-1}$ cm$^{-2}$                    \\ 
\hline
\multicolumn{8}{c}{Traditional analysis } & C-value/d.o.f\\
\hline
PL   &  $0.22$                 &  $3.1_{-0.4}^{+0.5}$ & $2.4\pm0.6$ & ... & ... & $4.3\pm 1.1$ & $20\pm5$& 201.56/212 \\
BB    &  $0.22$                 & ... & ... & $0.22_{-0.04}^{+0.05}$ & $73_{-30}^{+49}$ & $3.1\pm 0.7$ & $7.4\pm 1.8$& 197.88/212\\
\hline
\multicolumn{8}{c}{Maximum likelihood analysis } & $\chi_\nu^2/\mathrm{d.o.f}$ \\
\hline
PL             & 0.22  & $3.6_{-0.4}^{+0.5}$ & $2.0\pm0.5$   & ... & ... & $3.1\pm0.9$ & $26_{-10}^{+16}$& 1.40/9 \\
BB             & 0.22  & ... & ... & $0.21_{-0.03}^{+0.04}$ & $81_{-29}^{+46}$ & $2.9\pm0.7$ & $7.5_{-2.0}^{+2.2}$& 0.40/9 \\
\hline
\end{tabular}
\tablenotetext{a}{$10^{-6}$ photons cm$^{-2}$ s$^{-1}$ keV$^{-1}$ at 1 keV.\\}
\tablenotetext{b}{Blackbody radius for an assumed distance of 1.36 kpc.\\}
\end{center}
\end{table*}


\section{Discussion}

\subsection{Size of the polar cap and multipoles}

The X-ray emission from  PSR J1154--6250 has the  typical   properties  of old radio pulsars, in particular the small 
thermal luminosity of 
$\sim4\times 10^{29}$~erg~s$^{-1}$.  
We can also safely assume that the hot spot is not formed due to presence of the strong toroidal crust-confined magnetic field since its size is well below 10--40 degrees of latitude of the NS surface \citep{2008A&A...486..255A} which would correspond to a spot size of 1.7--7~km. The only plausible mechanism leading to the formation of such  tiny spots is surface heating by accelerated particles moving along the open field lines. 

The size of the polar cap heated by the in-falling accelerated particles depends strongly on the configuration of the magnetic field.
The polar cap radius
in the case of the pure dipole magnetic field is:
\begin{equation}
R_\mathrm{PC} = \sqrt{\frac{2\pi R^3_\mathrm{NS}}{cP} }
\label{e:Rpc}
\end{equation}
where $c$ is the speed of light.  
For PSR J1154--6250 this gives $R_\mathrm{PC}\approx 270$~m (for an assumed $R_{\rm NS}=10$ km), which is much larger than the value of the emitting radius derived from our blackbody fit. It is worth mentioning that the fit of spectra using NS hydrogen atmosphere model gives estimate of $R_\mathrm{BB}$ compatible with $R_\mathrm{PC}$. The discrepancy between $R_\mathrm{BB}$ and $R_\mathrm{PC}$ in the case of absorbed blackbody model could indicate the presence of high order magnetic   multipoles. These can increase  the curvature of the magnetic field lines, thus resulting  in a reduction of the polar cap size. This is not surprising since PSR J1154--6250  lays below the death line derived for pure dipolar magnetic field \citep{2000ApJ...531L.135Z, ChenRuderman1993,RS}:
\begin{equation}
\log \dot P = \frac{11}{4} \log P - 14.62,
\end{equation}
which results in $\log \dot P = -16.13$ for $P=0.28$~s, while the pulsar has $\log \dot P = -16.74$. The small scale magnetic field structure could   explain both the small size of the polar cap and the presence of radio emission. Another possibility  is that the small apparent size of the blackbody emitting region is simply due to a geometric  projection effect.

Of course, if the pulsar distance were a factor $\sim 4$ larger than assumed here, the 
polar cap size would correspond to the dipolar one.
Such a distance is allowed by the large uncertainties of the electron density model in rare cases \citep{2018ApJS..235...37A}. On the other hand, the older electron density model NE2001 \citep{2002astro.ph..7156C} provides a distance only 1.3 times larger.
The unknown equation of state for the NS matter hardly ever could contribute to this uncertainty since a range of 10.0~km up to 11.5~km \citep{2016ARA&A..54..401O} results only in $\approx $20\% variation in the polar cap radius.

\subsection{Comparison   with other pulsars}

It is known that old radio pulsars tend to have a higher efficiency of X-ray emission compared  to young   pulsars \citep{2012ApJ...749..146P}. The X-ray efficiency $\eta_X$ is the ratio of X-ray luminosity $L_X^\mathrm{nontherm}$ to $\dot E$:
\begin{equation}
\eta_X = \frac{4\pi D^2 F^\mathrm{unabs}_{0.2-10~\mathrm{keV}}}{\dot E} \approx 4.5\times 10^{-3}
\end{equation}
under the assumption of isotropic emission and using the flux obtained with the traditional analysis. 
If we interpret the X-ray emission of PSR J1154--6250 as non-thermal, we see that its efficiency is very similar to that of other old pulsars, e.g. comparable with that of PSR J0108--1431 \citep{earlier_old_pulsar}  and 2-3 orders of magnitude higher than the efficiency of young radio pulsars, see e.g. Fig.~5 in \cite{2012ApJ...749..146P}. In particular this efficiency is higher than  $2\times 10^{-3}$ which is expected from the relation by \cite{2002A&A...387..993P}. It is worth noting that relation of \cite{2002A&A...387..993P} was derived for X-ray luminosities in 2--10~keV energy band where we see virtually no emission from our object.

The power-law exponent found in our analysis is $\Gamma \approx 3.3$. This value is well within range described by \cite{2012ApJ...749..146P}. The value $\Gamma=3.1$ was found for another old pulsar PSR J2043+2740 ($\tau=1.2$~Myr) by \cite{2004ApJ...615..908B}.

\section{Conclusions}
We have detected for the first time the X-ray emission from the old radio pulsar PSR J1154--6250 using the \textit{XMM-Newton} observatory. 
The faintness of the source does not allow to clearly discriminate between emission of thermal or non-thermal origin.

In any case the spectrum is rather soft: a power-law with photon index $\Gamma\approx 3.3$ or a blackbody with temperature $kT=0.21$ keV. If the emission is non-thermal, the implied X-ray efficiency is similar to that of other pulsars of similar age. If it is instead of thermal origin, the low flux implies emission from a small region of radius $\approx 80$ m, most likely a polar cap reheated by backflowing magnetospheric particles.

All these properties of  PSR J1154--6250 are similar to those of other old radio pulsars. Therefore, we conclude that PSR J1154--6250 is old and its projection on the OB association Cru OB1 is most likely  a chance coincidence.

\acknowledgments
AI  would like to acknowledge support from the Israel science foundation (ISF) I\textendash CORE grant 1829/12. Work of SP was supported by the RSF grant 14-12-00146. This research was also supported by the grant 14.W03.31.0021 of the  Ministry of Education and Science of the Russian Federation (ST and AM). JE acknowledges partial support from NWO VIDI award A.2320.0076. We thank an anonymous referee for comments which helped to improve the manuscript.

This work is based on observations obtained with XMM-Newton, an European Space Agency (ESA) science mission with instruments and contributions directly funded by ESA Member States and NASA.
We used data from the  ESA mission
{\it Gaia} (\url{https://www.cosmos.esa.int/gaia}), processed by the {\it Gaia}
Data Processing and Analysis Consortium (DPAC,
\url{https://www.cosmos.esa.int/web/gaia/dpac/consortium}). Funding for the DPAC
has been provided by national institutions, in particular the institutions
participating in the {\it Gaia} Multilateral Agreement.
This research made use of the cross-match service provided by CDS, Strasbourg.	

The Digitized Sky Surveys were produced at the Space Telescope Science Institute under U.S. Government grant NAG W-2166. The images of these surveys are based on photographic data obtained using the Oschin Schmidt Telescope on Palomar Mountain and the UK Schmidt Telescope. The plates were processed into the present compressed digital form with the permission of these institutions.

\facility{\textit{XMM-Newton} }

\software{
astropy \citep{2013A&A...558A..33A},
SAS (v16.1.0 \citealt{2004ASPC..314..759G}), 
XSPEC (v12.9.1 \citealt{1996ASPC..101...17A})}

\bibliographystyle{aasjournal} 
\bibliography{old_pulsar}

\appendix

\section{Improving of astrometry for the X-ray source}

To improve the accuracy on the position of the detected X-ray source we corrected the \textit{XMM-Newton} astrometry using the second Gaia data release \citep{2016A&A...595A...1G,2018arXiv180409365G}. We first selected 26 point-like X-ray sources detected in the pn image with the highest significance by means of the \texttt{edetect\_chain} (excluding  the source under study). This list of X-ray sources was then cross-matched with the second Gaia data release based on coordinates coincidence within a circle of $4''$. From the resulting list we then selected only the objects with a ratio of X-ray to optical flux $\log f_x / f_G < -0.5$. Objects with $\log f_x / f_G > -0.5$ are expected to be galaxies for which the second Gaia data release does not provide precise position. Since the Gaia G band is quite similar to the V band of the Johnson system, especially if $(B-V)\approx 0$ \citep{2010A&A...523A..48J}, we computed $f_x / f_G$ as $\log f_x/f_G = \log f_x + 0.4 g  + 5.37$.
We optimized the residuals between the X-ray pn and the Gaia coordinates   
allowing rotation and translation of the X-ray frame using \texttt{astropy} \citep{2013A&A...558A..33A}. After a first optimization, we removed the sources with residual distances larger than $3''$ and optimized again. This optimization procedure for 17 sources decreased the mean displacement from the initial value of $1\farcs 33$ to $\sigma_\mathrm{fit}=1\farcs 21$. From the fit we conclude that the relative astrometric uncertainty is $3\sigma_\mathrm{fit}=3\farcs 63$. Since the Gaia  astrometric accuracy for stars with $g < 18^\mathrm{m}$ is $<0.07$~mas,  the main source of uncertainty is the statistical error on the X-ray positions.



\end{document}